\documentclass{article}
\usepackage{spconf,amsmath}
\usepackage{amsthm}
\usepackage{amsfonts}
\usepackage{amssymb}
\usepackage{mathrsfs}
\usepackage{mathtools}
\usepackage[english]{babel}
\usepackage{float}
\usepackage[toc,page]{appendix}
\usepackage[pdftex]{graphicx}
\usepackage{epstopdf}
\usepackage{amsmath}
\usepackage{xcolor}
\usepackage{url}
\usepackage{multirow}
\usepackage{graphicx}
\usepackage{IEEEtrantools}
\usepackage{bm}
\usepackage{balance}
\usepackage{microtype}
\usepackage[caption=false,font=footnotesize]{subfig}
\usepackage{cite}
\usepackage{comment}
\allowdisplaybreaks

\numberwithin{equation}{section}

\newtheorem{theorem}{Theorem}[section]
\makeatletter
\newcommand{\settheoremtag}[1]{
  \let\oldthetheorem\thetheorem
  \renewcommand{\thetheorem}{#1}
  \g@addto@macro\endtheorem{
    \addtocounter{theorem}{-1}
    \global\let\thetheorem\oldthetheorem}
  }
\makeatother

\usepackage[switch]{lineno}
\usepackage[named]{algo}
\usepackage[noend]{algpseudocode}
\usepackage{algorithm}

\newcommand{\suml}{\sum_{\ell=1}^{L}}
\newcommand{\sump}{\sum_{p=1}^{P}}
\newcommand{\sumk}{\sum_{k=1}^{K}}
\newcommand{\sumq}{\sum_{q=1}^{Q}}

\newcommand{\bsym}{\boldsymbol}

\DeclareMathOperator*{\minimize}{minimize }

\newtheorem{proposition}[theorem]{Proposition}

\ninept

\newcommand{\dddag}{%
  \mathbin{\vbox{\offinterlineskip\ialign{%
    \hfil##\hfil\cr
    \small{$\dagger$}\cr
    \noalign{\kern-0.6ex}
    \small{$\ddagger$}\cr
}}}}

\title{
Multi-dimensional dual-blind deconvolution approach toward joint radar-communications} 
	
\name{Roman Jacome$^{\dag}$, Kumar Vijay Mishra$^{\ddag}$, Edwin Vargas$^{\dag}$, Brian M. Sadler$^{\ddag}$, and Henry Arguello$^{\dag}$}
\address{$^{\dag}$Universidad Industrial de Santander, Bucaramanga, Colombia, 680002\\ 
$^{\ddag}$United States DEVCOM Army Research Laboratory, Adelphi, MD 20783 USA
\thanks{This research was sponsored by the Army Research Office/Laboratory under Grant Number W911NF-21-1-0099, and the VIE project 9834 entitled ``Dual blind deconvolution for joint radar-communications processing''. K. V. M. acknowledges partial support from the National Academies of Sciences, Engineering, and Medicine via Army Research Laboratory Harry Diamond Distinguished Postdoctoral Fellowship. Research was sponsored by the Army Research Laboratory and was accomplished under Cooperative Agreement Number W911NF-21-2-0288. The views and conclusions contained in this document are those of the authors and should not be interpreted as representing the official policies, either expressed or implied, of the Army Research Laboratory or the U.S. Government. The U.S.Government is authorized to reproduce and distribute reprints for Government purposes notwithstanding any copyright notation herein.}
}

\begin{document}
\setlength{\abovedisplayskip}{3pt}
\setlength{\belowdisplayskip}{3pt}

\maketitle

\begin{abstract}
We consider a joint multiple-antenna radar-communications system in a co-existence scenario. Contrary to conventional applications, wherein at least the radar waveform and communications channel are known or estimated \textit{a priori}, we investigate the case when the channels and transmit signals of both systems are unknown. In radar applications, this problem arises in multistatic or passive systems, where transmit signal is not known. Similarly, highly dynamic vehicular or mobile communications may render prior estimates of wireless channel unhelpful. In particular, the radar signal reflected-off multiple targets is overlaid with the multi-carrier communications signal. In order to extract the unknown continuous-valued target parameters (range, Doppler velocity, and direction-of-arrival) and communications messages, we formulate the problem as a sparse dual-blind deconvolution and solve it using atomic norm minimization. Numerical experiments validate our proposed approach and show that precise estimation of continuous-valued channel parameters, radar waveform, and communications messages is possible up to scaling ambiguities.  
\end{abstract}

\section{Introduction}
\label{sec:intro}

The increasingly limited spectrum for radar and communications applications has led to the development of joint radar-communications (JRC) systems. This emerging spectrum-sharing paradigm also has advantages of low cost, compact size, and less power consumption \cite{mishra2019toward,duggal2020doppler,elbir2021terahertz,elbir2022the}. Broadly, the following JRC modalities have emerged: co-design \cite{liu2020co}, cooperation \cite{bicua2018radar}, and co-existence \cite{wu2021resource}. In spectral co-design, a common transmit waveform and hardware units are envisaged to achieve an optimal spectrum usage. The cooperation model requires information from one system to aid the objectives of the other thereby leading to sensing-assisted communications and communications-assisted sensing applications. In spectral co-existence, radar and communications transmit and access the channel independently and focus on mitigating the mutual interference at the receiver. This scenario presents more difficult challenges in separating the overlaid radar and communications signals at the receiver. In this paper, we focus on spectral coexistence problem.

Conventionally, the transmit waveform of radar is known at the receiver and this knowledge is useful in extracting the unknown target parameters. In wireless communications, the channel estimates are available to the receiver, whose goal is to estimate the unknown transmitted messages. However, in certain radar and communications applications both signals and channels are unknown to the receiver. For instance, passive \cite{sedighi2021localization} and multistatic \cite{dokhanchi2019mmwave} sensing for low-cost and efficient covert operations may not have knowledge of the transmitted waveform \cite{kuschel2019tutorial}. In mobile radio \cite{neskovic2000modern} and vehicular networks \cite{olariu2009vehicular}, the channel is highly dynamic and its prior estimates may be outdated. Therefore, a general spectral coexistence scenario comprises of a common receiver \cite{vouras2022overview}, wherein both radar and communications channels and their respective transmit signals are unknown. 

In our previous work \cite{vargas2022joint}, we modeled the extraction of all four of these quantities, i.e. radar and communications channels and signals, as a \textit{dual-blind deconvolution} (DBD), wherein the observation is a sum of two convolutions and all four signals being convolved need to be estimated. This formulation is related to (single-)blind deconvolution (BD), a longstanding problem that occurs in a variety of engineering and scientific applications \cite{jefferies1993restoration,ayers1988iterative,abed1997blind}. The DBD problem in \cite{vargas2022joint} employed a single antenna and did not estimate direction-of-arrival (DoA) for either radar targets or communications signals.

In this paper, we study DBD for spectral coexistence when the receiver employs a uniform linear array (ULA) antenna. As a result, there are three continuous-valued target parameters (range, Doppler velocity and DoA), communications messages and communications DoA need to be estimated. We solve this three-dimensional (3-D) DBD by exploiting the sparsity of both radar and communications channels and formulating the problem as an atomic norm minimization (ANM) \cite{chandrasekaran2012convex,off_the_grid}. The ANM facilitates recovering continuous-valued parameters and has been previously leveraged in applications such as line spectrum denoising \cite{bhaskar2013atomic}, spectral super-resolution \cite{mishra2015spectral,xu2014precise}, multi measurement vector line spectrum estimation \cite{li2015off}, and DoA estimation \cite{chen2020new}. 
Among prior works, ANM was employed for 1-D BD in \cite{chi2016guaranteed} using a subspace representation of the modulating signal \cite{yang2016super}. Some studies have applied ANM to 2-D \cite{suliman2021mathematical} and 3-D \cite{suliman2019exact} blind super-resolution. In this paper, following our previous work \cite{vargas2022joint} that also exploited ANM to solve 2-D DBD, we cast the 3-D DBD as a multi-variate ANM. We then obtain the semidefinite program (SDP) of its dual problem by using properties of positive trigonometric polynomials.

The rest of the paper is organized as follows. In the next section, we present the coexistence system model of the 3-D DBD. Section \ref{sec:formulation} presents the proposed 3-D ANM formulation, its dual problem and SDP. Section \ref{sec:experiments} validates our proposed approach via several numerical experiments. We conclude in Section~\ref{sec:summ}. 

Throughout this paper, we reserve boldface lowercase, boldface uppercase, and calligraphic letters for vectors, matrices, and index sets, respectively. We denote the transpose, conjugate, Hermitian, and trace by $(\cdot)^T$, $(\cdot)^*$, $(\cdot)^H$, and $\text{Tr}(\cdot)$, respectively. The identity matrix of size $N\times N$ is $\mathbf{I}_N$. $||\cdot||_p$ is the $\ell_p$ norm. For notational convenience, the variables with subindex $r$ refer to the signals and parameters related to the radar system, while those with subindex $c$ refer to the communications system, we denote the $n-$th entry of a vector as $\mathbf{x}[n]$.
\section{System Model}
\begin{figure}[t]
    \centering
    \includegraphics[width=0.8\linewidth]{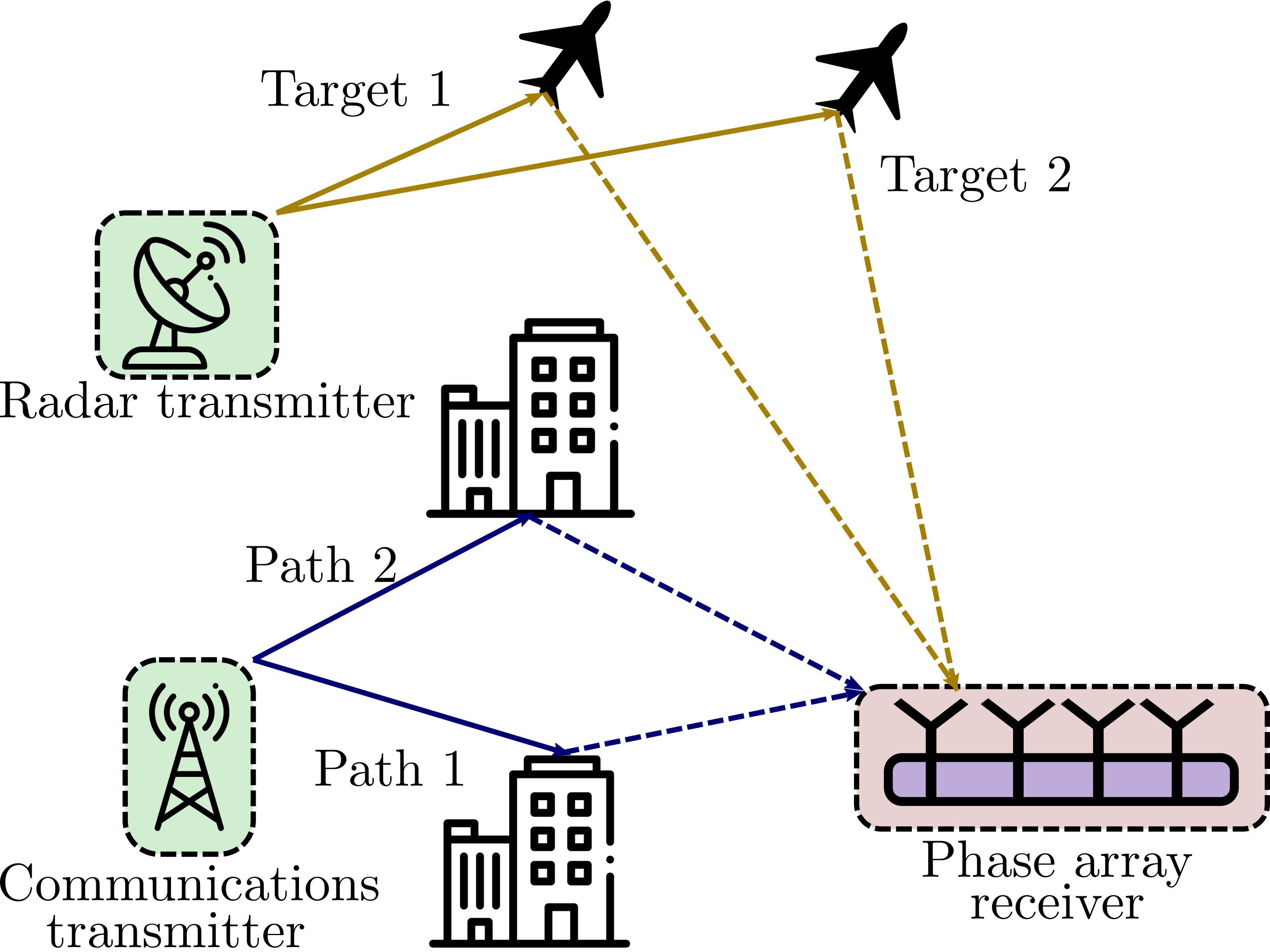}
    \caption{Independent radar and communications sources transmit toward multiple targets and through multiple paths respectively. The radar signal reflected-off the targets and communications signals are received by a ULA-based JRC receiver.}
    
    
    \label{fig:system}
\end{figure}

Consider a ULA-based receiver (Fig.~\ref{fig:system}) with $N_r$ antennas. The receiver admits overlaid radar and communications signals convolved by their respective channels. 
The transmit radar signal $x_r(t)=\sump s(t-pT)$ is a train of $P$ pulses $s(t)$ transmitted at a pulse repetition interval (PRI) $T$. The transmitted communications signal is a set of $P$ messages such that $x_c(t) = \sump v_p(t-pT)$, where $v(t)$ is a modulated orthogonal frequency-division multiplexing (OFDM) signal with $K$ modulating frequencies given by $v_p(t) =\sumk\mathbf{g}_p[k]e^{-\mathrm{j}2\pi k\Delta ft}$, where $\mathbf{g}_p [k]$ is the $p$-th message modulated by the $k$-th frequency. 

Consider $L$ radar targets, whose unknown parameters are encapsulated in $L\times 1$ vectors $\bsym{\alpha}_r, \bsym{\tau}_r, \bsym{\nu}_r$ and $\bsym{\beta}_r$ which contain the targets' complex reflectivity, time delays, Doppler velocities and DoA, respectively. The radar channel is 
\begin{align}
 \mathbf{h}_r(t) = \suml\bsym{\alpha}_r[\ell]\mathbf{b}(\bsym{\beta}_r[\ell])\delta(t-\bsym{\tau}_r[\ell])e^{-\mathrm{j}2\pi\bsym{\nu}_r[\ell] t},   
\end{align}
where $\mathbf{b}(\beta) = [1,e^{-\mathrm{j}2\pi \beta},\dots,e^{-\mathrm{j}2\pi (N_r-1)\beta}], \beta = \sin(\theta)/2$ is a steering vector, $\theta$ is the angle of arrival, and $\beta$ is DoA. Similarly, the communications channel is
\begin{align}
   \mathbf{h}_c(t) = \sumq\bsym{\alpha}_c[q]\mathbf{b}(\bsym{\beta}_c[q])\delta(t-\bsym{\tau}_c[q])e^{-\mathrm{j}2\pi\bsym{\nu}_c[q] t}. 
\end{align}
The received signal at the ULA receiver is
\begin{align}
    \mathbf{y}(t) =&[y_1(t),\dots,y_{N_r}(t)]^T\nonumber\\
    =&x_r(t)*\mathbf{h}_r(t)+x_c(t)*\mathbf{h}_c(t)\nonumber\\
    =&\sump \suml\bsym{\alpha}_r[\ell]\mathbf{b}(\bsym{\beta}_r[\ell])s(t-pT-\bsym{\tau}_r[\ell])e^{-\mathrm{j}2\pi\bsym{\nu}_r[\ell] t}+\nonumber\\& \sumk\sumq \bsym{\alpha}_c[q]\mathbf{g}_p[k]e^{-\mathrm{j}2\pi k\Delta f(t-pT-\bsym{\tau}_c[q])}\mathbf{b}(\bsym{\beta}_c[q])e^{-\mathrm{j}2\pi\bsym{\nu}_c[q] t}.\nonumber
\end{align}

Rewrite the received signal as $\mathbf{y}(t)= \sump \tilde{\mathbf{y}}_p(t)$. Our measurements are determined in terms of shifted signals $\mathbf{y}_p(t) = \hat{\mathbf{y}}_p(t + pT)$, such that the signals $\hat{\mathbf{y}}_p(t+ pT)$ are time-aligned with $\mathbf{y}_0(t)$. Therefore, the signal $\mathbf{y}_1(t)$ and the shifted signals $\mathbf{y}_p(t)$ contain the same set of parameters. We compute the continuous-time Fourier transform (CTFT) of $\mathbf{y}_p(t)$ in $f\in[-\frac{B}{2},\frac{B}{2}]$, with $p = 1,\dots, P$ and uniform sampling at $f_n= \frac{Bn}{M}=n\Delta f$, with $n = -N,\dots,N$, $M = 2N+1$. For the sake of simplicity, set $M = K$, i.e. samples in the frequency domain at the OFDM separation frequency $\Delta f$ \cite{zheng2017super};  however, this is not necessary for our recovery procedure. This produces the CTFT as
\begin{align}
\tilde{\mathbf{y}}_p(f_n) = &\suml\bsym{\alpha}_r[\ell]\tilde{s}(f_n)\mathbf{b}(\bsym{\beta}_r[\ell])e^{-\mathrm{j}2\pi(n\Delta f\bsym{\tau}_r[\ell] +\bsym{\nu}_r[\ell] pT)}+\nonumber\\& \sumq\mathbf{g}_p[n]\mathbf{b}(\bsym{\beta}_c[q])e^{-\mathrm{j}2\pi(n\Delta f\bsym{\tau}_c[q]+\bsym{\nu}_c[q] pT)},\label{ft_2}
\end{align}
where $\tilde{s}(f)$ is the Fourier transform of $s(t)$. 

We introduce the index sequence $m = 1,\dots,MP$ where $m = n+N+Np$. The vector $\mathbf{y} = [\tilde{\mathbf{y}}_0^T, \dots,  \tilde{\mathbf{y}}_{MP}^T]^T \in \mathbb{C}^{N_rMP}$ contains all samples for every receiver antenna and $ \tilde{\mathbf{y}}_m\in\mathbb{C}^{N_r}$. Normalize the parameters ${\bsym{\tau}}_r[\ell] = \frac{\bar{\bsym{\tau}}_r[\ell]}{T}$, $\bsym{\nu}_r[\ell] = \frac{\bar{\bsym{\nu}}_r[\ell]}{\Delta f}$, ${\bsym{\tau}}_c[q] = \frac{\bar{\bsym{\tau}}_c[q]}{T}$, $\bsym{\nu}_c[q] = \frac{\bar{\bsym{\nu}}_c[q]}{\Delta f}$, such that ${\bsym{\tau}}_r[\ell],{\bsym{\nu}}_r[\ell], {\bsym{\beta}}_r[\ell] {\bsym{\tau}}_c[q],{\bsym{\nu}}_c[q], {\bsym{\beta}}_c[q]  \in [0,1] $. Vectorize $\tilde{s}(f_n)$ as $\mathbf{s}[n] = \tilde{s}(f_n)$  and $\mathbf{g}[m] = \mathbf{g}_p[n]$. This yields 
\begin{align}
\tilde{\mathbf{y}}_m = &\suml\bsym{\alpha}_r[\ell]\mathbf{s}[n]\mathbf{b}(\bsym{\beta}_r[\ell])e^{-\mathrm{j}2\pi(n\bsym{\tau}_r[\ell] +\bsym{\nu}_r[\ell] p)}+\nonumber\\& \sumq\bsym{\alpha}_c[q]\mathbf{g}[m]\mathbf{b}(\bsym{\beta}_c[q])e^{-\mathrm{j}2\pi(n\bsym{\tau}_c[q]+\bsym{\nu}_c[q] p)}.\label{y_p}
\end{align}

Our goal is to estimate the set of radar and communications parameters $\bsym{\alpha}_r$, $\bsym{\tau}_r$, $\bsym{\nu}_r$, $\bsym{\beta}_r$,  $\bsym{\alpha}_c$, $\bsym{\tau}_c$, $\bsym{\nu}_c$, and $\bsym{\beta}_c$, when the radar pulses $\mathbf{s}$ and communications symbols $\mathbf{g}$ are also unknown. In this inverse problem, the number of unknowns is $O(3LMPN_r(L+Q))$ and, therefore, it is highly ill-posed. Our strategy to solve this problem is by assuming that $\mathbf{s}$ and $\mathbf{g}$ lie in a given low-dimensional subspace \cite{chi2016guaranteed,yang2016super}, i.e., $\mathbf{s}=\mathbf{Tv}$, $\mathbf{g}=\mathbf{Du}$, where $\mathbf{v}\in\mathbb{C}^{K}$ is the unknown coefficient vector of the radar waveform, $\mathbf{u} = [\mathbf{u}_1^T,\dots,\mathbf{u}_P^T]^T \in \mathbb{C}^{PK}$ is a vector containing the $P$ coefficient vectors of the communications messages such that $\mathbf{u}_p\in\mathbb{C}^{K}$, the matrices $\mathbf{T} \in \mathbb{C}^{M\times K}$ and $\mathbf{D}\in \mathbb{C}^{MP\times PK}$ are the known random transformation matrices. Moreover, $\mathbf{D}=\operatorname{blockdiag}(\mathbf{D}_1,\dots,\mathbf{D}_P), \mathbf{D}_p\in\mathbb{C}^{M\times K}$ such that

\begin{equation}
\mathbf{D} = 
\left[\begin{array}{cccc}
\mathbf{D}_1& \mathbf{0} & \hdots & \mathbf{0}   \\
\mathbf{0} & \mathbf{D}_2 &  \hdots & \mathbf{0}   \\
\mathbf{0}&\mathbf{0} & \small{\ddots} & \vdots  \\
\mathbf{0} & \mathbf{0} & \hdots  & \mathbf{D}_P
\end{array}\right] \nonumber,
\end{equation}

Thus, we rewrite the signal in \eqref{y_p} as
\begin{align}
\tilde{\mathbf{y}}_m = &\suml\bsym{\alpha}_r[\ell]\mathbf{t}_n^{H}\mathbf{v}\mathbf{b}(\bsym{\beta}_r[\ell])e^{-\mathrm{j}2\pi(n\bsym{\tau}_r[\ell] +\bsym{\nu}_r[\ell] p)}+\nonumber\\& \sumq\bsym{\alpha}_c[q]\mathbf{d}_m^T\mathbf{u}\mathbf{b}(\bsym{\beta}_c[q])e^{-\mathrm{j}2\pi(n\bsym{\tau}_c[q]+\bsym{\nu}_c[q] p)}.
\end{align}
Define the steering vector for the continuous-valued time-delay and Doppler modulation parameters $\mathbf{a(\tau,\nu)} = \big[e^{\mathrm{j}2\pi(\tau (-N)+\nu (1))},\dots$ $,e^{\mathrm{j}2\pi(\tau (N)+\nu(0))},\dots,e^{\mathrm{j}2\pi(\tau (N)+\nu(P))}\big] \in\mathbb{C}^{MP}$   and the vector $\mathbf{w}(\mathbf{r}) = \mathbf{a}(\tau,\nu)\otimes\mathbf{b}(\beta) \in \mathbb{C}^{N_r MP}$  with $\mathbf{r} = [\bsym{\tau}_r,\bsym{\nu}_r,\bsym{\beta}_r]$. The channel vectors become $\mathbf{h}_r = \suml\bsym{\alpha}_r[\ell]\mathbf{w(\mathbf{r}_\ell)}$ and $\mathbf{h}_c = \sumq\bsym{\alpha}_c[q]\mathbf{w}(\mathbf{c}_q)$, where $\mathbf{c} = [\bsym{\tau}_c,\bsym{\nu}_c,\bsym{\beta}_c]$. We express the full measurement vector as
\begin{equation}
    \mathbf{y}_j = \mathbf{h}_r^H\mathbf{e}_j\mathbf{t}_n\mathbf{u} + \mathbf{h}_c^H\mathbf{e}_j\mathbf{d}_m\mathbf{v},
\end{equation}
where the index $j=1,\dots,N_rMP$ follows the sequence $j=m+MPr, r=1,\dots,N_r$ and $\mathbf{e}_j$ is the $j$-th canonical vector of $\mathbb{R}^{N_rMP}$. 

Define the matrices $\mathbf{G}_j = \mathbf{e}_j\mathbf{t}_n^H \in \mathbb{C}^{N_rMP\times K}, \mathbf{A}_j = \mathbf{e}_j\mathbf{d}_m^H \in \mathbb{C}^{N_rMP\times PK}$ as random sensing matrices. Denote $\mathbf{X}_r = \mathbf{u}\mathbf{h}_r^H \in \mathbb{C}^{K\times N_rMP}, \mathbf{X}_c = \mathbf{v}\mathbf{h}_c^H$ as rank-one matrices that contain the unknown variables (channel parameters and signal coefficients). The measurement vector is a linear transformation of $\mathbf{X}_r$ and $\mathbf{X}_c$  
\begin{equation}
    \mathbf{y} = \mathcal{B}_r(\mathbf{X}_r)+\mathcal{B}_c(\mathbf{X}_c),
\end{equation}
where the linear operator  $\mathcal{B}_r:\mathbb{C}^{K\times N_rMP}\rightarrow \mathbb{C}^{N_rMP}, \mathcal{B}_c:\mathbb{C}^{PK\times N_rMP}\rightarrow \mathbb{C}^{N_rMP}$ are defined as $\mathcal{B}_r(\mathbf{X}_r)[j] = \operatorname{Tr}(\mathbf{G}_j\mathbf{X}_r)$ and $ \mathcal{B}_c(\mathbf{X}_c)[j] = \operatorname{Tr}(\mathbf{A}_j\mathbf{X}_c)$. 
\section{Multi-Dimensional DBD}
\label{sec:formulation}
The radar and communications channels are characterized by a few continuous-valued parameters $L+Q\ll MPN_r$. Leveraging the sparse nature of these channels, we use ANM framework \cite{off_the_grid} for super-resolved estimations of continuous-valued channel parameters. 
For the overlaid radar-communications signal, we formulate the parameter recovery as the minimization of two atomic norms, each corresponding to the radar and communications signal trails. Define the sets of atoms for the radar and communications signals as, respectively,
\begin{align}
    &\mathcal{A}_r = \Big\{\mathbf{u}\mathbf{w}(\mathbf{r})^H: \mathbf{r}\in[0,1]^3,||\mathbf{u}||_2 = 1   \Big\}
    \\&\mathcal{A}_c = \Big\{\mathbf{v}\mathbf{w}(\mathbf{c})^H: \mathbf{c}\in[0,1]^3,||\mathbf{v}||_2 = 1   \Big\}.
    \label{eq:atomic_sets}
\end{align}
The corresponding atomic norms are
\begin{align}
    &||\mathbf{X}_r||_{\mathcal{A}_r} = \inf_{\stackrel{\bsym{\alpha}_r[\ell] \in \mathbb{C}, \boldsymbol{r}_\ell \in [0,1]^3}{||\mathbf{u}||_2 = 1}} \Bigg\{\sum_\ell \vert\bsym{\alpha}_r[\ell]\vert \Big| \mathbf{Z}_r = \sum_\ell \bsym{\alpha}_r[\ell]\mathbf{u}\mathbf{w}(\mathbf{r}_\ell)^H\Bigg\}\nonumber\\
    &||\mathbf{X}_c||_{\mathcal{A}_c} = \inf_{\stackrel{\bsym{\alpha}_c[q]\in \mathbb{C}, \boldsymbol{c}_q \in [0,1]^3}{||\mathbf{v}||_2 = 1}} \Bigg\{\sum_q \vert\bsym{\alpha}_c[q]\vert\Big| \mathbf{Z}_c = \sum_q \bsym{\alpha}_c[q]\mathbf{v}\mathbf{w}(\mathbf{c}_q)^H\Bigg\}\nonumber.
\end{align}
Consequently, our proposed ANM problem is
\begin{align}
    &\minimize_{\mathbf{X}_r,\mathbf{X}_c} ||\mathbf{X}_r||_{\mathcal{A}_r} +||\mathbf{X}_c||_{\mathcal{A}_c} 
    \;\text{subject to }     \mathbf{y} = \mathcal{B}_r(\mathbf{X}_r) + \mathcal{B}_c(\mathbf{X}_c).
    \label{eq:primal_problem}
\end{align}

In order to formulate the SDP of the above-mentioned ANM, we employed the dual optimization problem for the DBD developed in \cite{vargas2022joint} which results in 
\begin{align}
    &\underset{\mathbf{q}}{\textrm{maximize}}\langle\mathbf{q,y}\rangle_{\mathbb{R}}
    \text{subject to } \Vert\mathcal{B}_r^\star(\mathbf{q})\Vert^\star_{\mathcal{A}_r}\leq1, 
    \Vert \mathcal{B}_c^\star(\mathbf{q})\Vert^\star_{\mathcal{A}_c}\leq1, 
\label{eq:dual_problem_op}    
\end{align}
where $\mathcal{B}_r^\star: \mathbb{C}^{N_rMP}\rightarrow\mathbb{C}^{K \times N_rMP}$ and $\mathcal{B}_c^\star: \mathbb{C}^{N_rMP}\rightarrow\mathbb{C}^{PK \times N_rMP}$  are adjoint operators of $\mathcal{B}_r$ and $\mathcal{B}_c$, respectively, and defined as  $\mathcal{B}_r^\star(\boldsymbol{q}) = \sum_{r=1}^{N_r}\sump\sum_{n=-N}^{N}\mathbf{q}[j]\mathbf{G}_j^H$ and $\mathcal{B}_c^\star(\mathbf{q}) = \sum_{r=1}^{N_r}\sump\sum_{n=-N}^{N}\mathbf{q}[j]\mathbf{A}_j^H $. In order to formulate SDP, we use the following vector-valued positive trigonometric polynomials
\begin{align}
    &\mathbf{f}_r(\mathbf{r}) =  \sum_{r=1}^{N_r}\sump\sum_{n=-N}^{N}\mathbf{q}[j]\mathbf{G}_j^H\mathbf{w}(\mathbf{r}) \in \mathbb{C}^{K},
    \label{eq:poly_r}\\
    &\mathbf{f}_c(\mathbf{c})=\sum_{r=1}^{N_r}\sump\sum_{n=-N}^{N}\mathbf{q}[j]\mathbf{A}_j^H\mathbf{w}(\mathbf{c}) \in \mathbb{C}^{KP},
    \label{eq:poly_c}
\end{align}
each of which is parameterized by positive definite matrices \cite{dumitrescu2007positive}. 

Using the Bounded Real Lemma \cite{dumitrescu2007positive}, we convert the constraints on \eqref{eq:dual_problem_op} to linear matrix inequalities. The optimization problem in \eqref{eq:dual_problem_op} is equivalent to the SDP 
\begin{align}
    &\underset{\mathbf{q,Q}}{\textrm{maximize}}\quad \langle\mathbf{q,y}\rangle_{\mathbb{R}}\nonumber\\
    &\text{subject to }\mathbf{Q}\succeq 0,\nonumber\\&\hphantom{\text{subject to }}  
    \begin{bmatrix}
        \mathbf{Q} & \hat{\mathbf{Q}}_r^H \\
        \hat{\mathbf{Q}}_r & \mathbf{I}_K
        \end{bmatrix}
    \succeq0,\nonumber\\&\hphantom{\text{subject to }} 
    \begin{bmatrix}
        \mathbf{Q} & \hat{\mathbf{Q}}_c^H \\
        \hat{\mathbf{Q}}_c & \mathbf{I}_{KP} 
        \end{bmatrix}\succeq 0,
    \nonumber\\&\hphantom{\text{subject to }}
    \text{Tr}(\boldsymbol{\Theta}_\mathbf{n}\mathbf{Q}) = \delta_{\mathbf{n}},\label{dual_opt}
\end{align}
where $\hat{\mathbf{Q}}_r =  \sum_{r=1}^{N_r}\sump\sum_{n=-N}^{N}\mathbf{q}[j]\mathbf{G}_j^H\in \mathbb{C}^{N_rMP\times K}$ and $\hat{\mathbf{Q}}_c =\sum_{r=1}^{N_r}\sump\sum_{n=-N}^{N}\mathbf{q}[j]\mathbf{A}_j^H\in \mathbb{C}^{N_rMP\times PK}$ are the coefficients of 3-D trigonometric polynomials, the matrix 
$\boldsymbol{\Theta}_\mathbf{n} = \boldsymbol{\Theta}_{n_3} \otimes \boldsymbol{\Theta}_{n_2} \otimes \boldsymbol{\Theta}_{n_1}$, where $\bsym{\Theta}_n$ is the Toeplitz matrix with ones in the $n$-th diagonal with $0<n_1<m_1$, $-m_2<n_2<m_2$ and $-m_3<n_3<m_3$ . Here, we define $m_1 = P-1$, $m_2=N-1$ and $m_3 = N_r-1$. Finally, $\delta_\mathbf{n} = 1 $ if $\mathbf{n} = [0,0,0]$ and $0$ otherwise. 
This SDP formulation is solved by employing off-the-shelf solvers. 
The following proposition states the conditions for exact recovery of the radar and communications channels parameters.  
\begin{figure*}[t]
    \centering
    \includegraphics[width=0.9\linewidth]{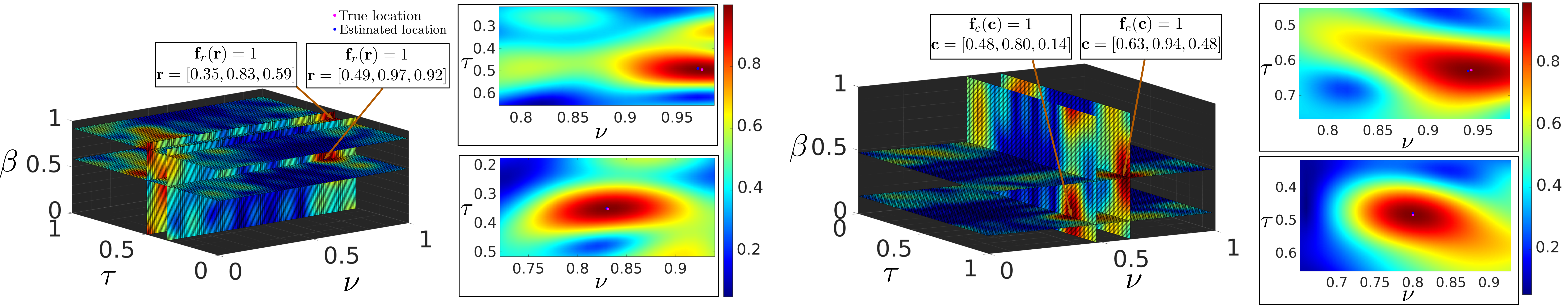}
    \caption{Radar and communications channel parameter estimation in the 3-D vector-valued positive trigonometric polynomial. }
    \label{fig:results_dual}
\end{figure*}
\begin{proposition}
 Denote $\mathcal{R} = \{\mathbf{r}_\ell\}_{\ell=0}^{L-1}$ and $\mathcal{C} = \{\mathbf{c}_q\}_{q=0}^{Q-1}$. The solutions of \eqref{eq:primal_problem} are $\hat{\mathbf{Z}}_r$ and $\hat{\mathbf{Z}}_c$. Then, $\hat{\mathbf{Z}}_r={\mathbf{Z}}_r$ and $\hat{\mathbf{Z}}_c={\mathbf{Z}}_c$ are the optimal solutions of \eqref{eq:primal_problem} if there exist two 3-D trigonometric polynomials with complex coefficients $\mathbf{q}$ such that
\begin{align}
    \mathbf{f}_r(\mathbf{r}_\ell) &= \mathrm{sign}( \bsym{\alpha}_r[\ell]) \mathbf{u} \hspace{1em} \text{if} \hspace{1em} \forall \mathbf{r}_\ell \in \mathcal{R}, \label{eq:cert_1}\\
    \mathbf{f}_c(\mathbf{c}_q) &= \mathrm{sign}( \bsym{\alpha}_c[q]) \mathbf{v} \hspace{1em} \text{if} \hspace{1em} \forall\mathbf{c}_q \in \mathcal{C},  \label{eq:cert_2}\\
    \Vert \mathbf{f}_r(\mathbf{r}) \Vert_2^2 &< 1 \hspace{1em } \forall \mathbf{r} \in [0,1]^3 \setminus \mathcal{R},  \label{eq:cert_3}\\
    \Vert \mathbf{f}_c(\mathbf{c}) \Vert_2^2  &< 1 \hspace{1em }  \forall \mathbf{c} \in [0,1]^3 \setminus \mathcal{C},\label{eq:cert_4}
\end{align}
where $\operatorname{sign}(c) = \frac{c}{|c|}$ .
\begin{proof}
The variable $\mathbf{q}$ is dual feasible. \par\noindent\small
\begin{flalign}
    &\langle\mathbf{q,y}\rangle_{\mathbb{R}} = \langle\mathcal{B}_r^*(\mathbf{q}),\mathbf{X}_r\rangle_{\mathbb{R}}+ \langle\mathcal{B}_c^*(\mathbf{q}),\mathbf{X}_c\rangle_{\mathbb{R}}\nonumber\\
    & = \suml \bsym{\alpha}_r[\ell]^* \langle\mathcal{B}_r^*(\mathbf{q}),\mathbf{u}\mathbf{w}(\mathbf{r}_\ell)^H\rangle_{\mathbb{R}} + \sumq \bsym{\alpha}_c[q]^* \langle\mathcal{B}_c^*(\mathbf{q}),\mathbf{v}\mathbf{w}(\mathbf{c}_q)^H\rangle_{\mathbb{R}}\nonumber \\
    &= \suml \bsym{\alpha}_r[\ell]^* \langle\mathbf{f}_r(\mathbf{r}_\ell),\mathbf{u}\rangle_{\mathbb{R}} + \sumq \bsym{\alpha}_c[q]^* \langle\mathbf{f}_c(\mathbf{c}_q),\mathbf{v}\rangle_{\mathbb{R}}\nonumber\\
    & =\suml\bsym{\alpha}_r[\ell]^*\textrm{sign}( \bsym{\alpha}_r[\ell]) + \sumq \bsym{\alpha}_c[q]^*\textrm{sign}( \bsym{\alpha}_c[q])\nonumber\\
    &= \suml |\bsym{\alpha}_r[\ell]| + \sumq |\bsym{\alpha}_c[q]|\geq ||\mathbf{X}_r||_{\mathcal{A}_r} + ||\mathbf{X}_c||_{\mathcal{A}_c}.\label{eq:lower_bound_dual}
\end{flalign}\normalsize
On the other hand, it follows from H\"{o}lder inequality that
\begin{align}
\langle\mathbf{q,y}\rangle_{\mathbb{R}}&=\langle\mathcal{B}_r^*(\mathbf{q}),\mathbf{X}_r\rangle_{\mathbb{R}}+ \langle\mathcal{B}_c^*(\mathbf{q}),\mathbf{X}_c\rangle_{\mathbb{R}}\\
&\leq ||\mathcal{B}_r^*(\mathbf{q})||_{\mathcal{A}_r}^*||\mathbf{X}_r||_{\mathcal{A}_r} + ||\mathcal{B}_c^*(\mathbf{q})||_{\mathcal{A}_c}^*||\mathbf{X}_c||_{\mathcal{A}_c}\\
&\leq ||\mathbf{X}_r||_{\mathcal{A}_r}+||\mathbf{X}_c||_{\mathcal{A}_c},
\label{eq:upper_bound_dual}
\end{align}
where the first inequality is due to Cauchy-Schwarz inequality and the last inequality follows from \eqref{eq:cert_1}, \eqref{eq:cert_2}, \eqref{eq:cert_3}, and \eqref{eq:cert_4}. Therefore, based on \eqref{eq:lower_bound_dual} and \eqref{eq:upper_bound_dual}, we conclude that $\langle\mathbf{q,y}\rangle_{\mathbb{R}}=||\mathbf{X}_r||_{\mathcal{A}_r}+||\mathbf{X}_c||_{\mathcal{A}_c}$ showing that the pair $(\mathbf{X}_r,\mathbf{X}_c)$ is primal optimal and, from strong duality, $\mathbf{q}$ is dual optimal.
\end{proof}
\end{proposition}

Computing the dual polynomials in \eqref{eq:poly_r} and \eqref{eq:poly_c} yields the estimated channel parameters $\{\hat{\mathbf{r}}_\ell\}_{\ell=1}^{L}, \{\hat{\mathbf{c}}_q\}_{q=1}^{Q}$. The radar and communications coefficients vectors $\mathbf{v} $ and $\mathbf{u}$ are then estimated by solving an over-determined linear system of equations. Define the matrices $\mathbf{W}_r \in \mathbb{C}^{MPN_r\times LJ}$ and $\mathbf{W}_c \in \mathbb{C}^{MPN_r\times PQJ}$ as 
\begin{equation*}
\mathbf{W}_r = 
\left[\begin{array}{ccc}
\mathbf{w}\left(\hat{\mathbf{r}}_{1}\right)^{H} \mathbf{A}_{1} & \ldots & \mathbf{w}\left(\hat{\mathbf{r}}_{L}\right)^{H}  \mathbf{A}_{1} \\
\vdots & \ddots & \vdots \\
\mathbf{w}\left(\hat{\mathbf{r}}_{0}\right)^{H} \mathbf{A}_{MPN_r} & \ldots & \mathbf{w}\left(\hat{\mathbf{r}}_{L}\right)^{H}  \mathbf{A}_{MPN_r}
\end{array}\right],
\end{equation*}
and
\begin{equation*}
\mathbf{W}_c = 
\left[\begin{array}{ccc}
\mathbf{w}\left(\hat{\mathbf{c}}_{1}\right)^{H} \mathbf{G}_{1} & \ldots & \mathbf{w}\left(\hat{\mathbf{c}}_{Q}\right)^{H}  \mathbf{G}_{1} \\
\vdots & \ddots & \vdots \\
\mathbf{w}\left(\hat{\mathbf{c}}_{1}\right)^{H} \mathbf{G}_{MPN_r} & \ldots & \mathbf{w}\left(\hat{\mathbf{c}}_{Q}\right)^{H}  \mathbf{G}_{MPN_r}
\end{array}\right].
\end{equation*}

Denote the vector containing the  desired coefficient vectors as $\mathbf{z} = \left[\bsym{\alpha}_c[1]\mathbf{u},\dots,\bsym{\alpha}_r[L]\mathbf{u}^T,\bsym{\alpha}_c[1]\mathbf{v}^T,\dots, \bsym{\alpha}_c[Q]\mathbf{v}^T\right]^T$ and define the matrix $\mathbf{W} = [\mathbf{W}_r,\mathbf{W}_c]$. The coefficient vector is then recovered (up to a scaling factor) by solving
$\mathbf{W}\mathbf{z} = \mathbf{y}$, through, say, least-squares. It only requires linear independence of columns of the matrix $\mathbf{W}$ because the matrices depends on the values of steering vectors $\mathbf{w}(\hat{\mathbf{r}}_\ell)$,  $\mathbf{w}(\hat{\mathbf{c}}_q)$. When the parameter set $\mathbf{r}$ achieves a minimum separation \cite{heckel2016superMIMO}, i.e. $\vert \beta_i - \beta_k\vert\geq\frac{5}{N_r},\vert \nu_i - \nu_k\vert\geq\frac{5}{P},\vert \tau_i - \tau_k\vert\geq\frac{5}{M}, \text{ for all } k\neq i$, the system matrix has full column rank. 
\begin{figure}[t!]
    \centering
\includegraphics[width=\linewidth]{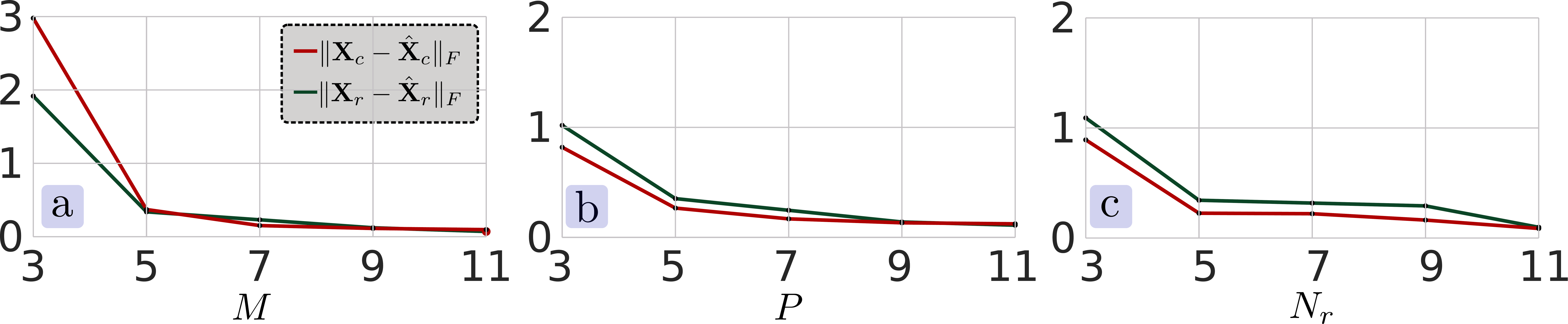}\qquad
    \caption{Statistical performance of ANM-based DBD in terms of recovery error when (a) number of samples $M$, (b) number of pulses/messages $P$ and (c) antennas $N_r$ are varied.}
    \label{fig:stats}
\end{figure}
\section{Numerical Experiments}
\label{sec:experiments}
To evaluate the proposed method, we considered a scenario with $M = 9$, $P=9$, $N_r = 3$, $Q=L=2$ and $K=3$. The delay, Doppler, and DoA parameters were drawn from a random uniform distribution, which results in $\mathbf{r}_1 = [0.352 0.831,0.585]$, $\mathbf{r}_2 = [0.495,0.974,0.919]$, $\mathbf{c}_1=[0.485, 0.800,0.142]$, and $\mathbf{c}_2=[0.628,0.943,0.475]$. The columns of the transformation matrices $\mathbf{T}$ and $\mathbf{D}_p$ were generated following the distribution described in \cite{chi2016guaranteed}, i.e. $\mathbf{t}_n= [1, e^{\mathrm{j}2\pi \sigma_n}, \dots, e^{\mathrm{j}2\pi(K-1) \sigma_n}]$, where $\sigma_n \sim \mathcal{N}(0,1)$. The parameters $\bsym{\alpha}_r$ and $\bsym{\alpha}_c$ were drawn from a normal distribution with $|\bsym{\alpha}_r[\ell]|=|\bsym{\alpha}_c[q]|=1$. The coefficient vectors $\mathbf{u}, \mathbf{v}$ were generated from a normal random distribution and normalized $\Vert\mathbf{u}\Vert=\Vert\mathbf{v}\Vert=1$. We 
used the CVX SDPT3 solver \cite{grant2009cvx}. 

The solution to the dual problem yields the dual trigonometric polynomials, which we computed on discrete 3-D time-delay, Doppler, and DoA domains with a sampling step of $1e-3$. The resulting dual polynomials are shown in Fig. \ref{fig:results_dual}, where slices of the 3-D polynomial at the ground truth position on the $\beta$ and $\tau$ dimension are displayed. The polynomials$\Vert\mathbf{f}_r(\mathbf{r})\Vert$ and  $\Vert\mathbf{f}_c(\mathbf{c})\Vert$ are unity at the locations corresponding to the targets/paths. The plots are accompanied by 2-D slices in the $\tau$ and $\nu$ planes at the ground truth value of $\beta$. 

Next, we studied the statistical performance of the method by varying the number of samples $M$, number of pulses/messages $P$ and receivers $N_r$. We ran 40 realizations for each experiment and computed the mean of the Frobenious norm $\Vert\mathbf{X}_r-\hat{\mathbf{X}}_r\Vert_F$ and $\Vert\mathbf{X}_c-\hat{\mathbf{X}}_c\Vert_F$. Fig \ref{fig:stats}(a) shows the performance with varying the number of samples $M$ while keeping $P=5$ and $N_r=3$ fixed. Fig \ref{fig:stats}(b) illustrates the same by varying the number pulses/messages $P$ for fixed $M=5, N_r=3$. Finally, for $M=5$ and $P=3$, Fig. \ref{fig:stats}(c) plots the recovery error with changes in the number of antennas $N_r$.
\section{Summary}
\label{sec:summ}
We proposed a 3-D DBD approach for ULA-based JRC receiver. The channels of both radar and communications were modeled as sparse signals that encapsulated time-delays, Doppler velocities and DoA parameters. We minimized a sum of atomic norms to estimate these continuous-valued parameters. Utilizing the theories of positive trigonometric polynomials, we obtained the SDP of the dual problem as well as performance guarantees. The results show perfect recovery with sufficient number of samples as predicted by our analytical result.
\bibliographystyle{IEEEtran}
\bibliography{biblio.bib}

\end{document}